# Characterization of Shapiro steps in the presence of a 4π-periodic Josephson current

Jinho Park, Yong-Bin Choi, Gil-Ho Lee, and Hu-Jong Lee

*Pohang University of Science and Technology, Pohang, Korea*

## Abstract

The Majorana zero-energy modes (MZMs) residing at the boundary of topological superconductors have attracted a great deal of interest recently, as they provide a platform to explore fundamental physics such as non-Abelian statistics, as well as fault-tolerant quantum computation. Period doubling of Shapiro steps in a Josephson junction under microwave irradiation has been regarded as strong evidence for the emergence of the MZMs at the junction edges. However, questions remain as to how the Shapiro steps respond to the presence of a 4π-periodic Josephson current. In this study, we investigated the characteristic features of Shapiro steps with respect to the ratio ($\alpha$) of the 4π-periodic current to the topologically trivial 2π-periodic one, as well as the reduced microwave frequency ($\Omega$) and McCumber parameter ($\beta$) of the junction. Our analysis reproduced Shapiro steps similar to those observed experimentally for specific parameter sets of $\alpha$, $\Omega$ ($\lesssim 0.1$), and $\beta$ ($\gtrsim 1.0$). Full suppression of the first lobe of the $n=1$ step guarantees the presence of a 4π-periodic Josephson current. In addition, we discuss the range of $\Omega$ and $\beta$ needed for full suppression of the first lobe of the $n=1$ step, even for small $\alpha$ ($< 0.1$). To observe "period-doubled Shapiro steps", even with a small $\alpha$, the junction should have a large $I_c R_N$ product and sufficiently large junction capacitance.

## I. INTRODUCTION

The Majorana fermion is a particle that acts as its own antiparticle, satisfying $\gamma^+ = \gamma$, which was introduced by Ettore Majorana as a solution to the relativistic Dirac equation [1]. Although it is uncertain whether Majorana fermions exist in nature, it is predicted that quasiparticles in Majorana fermionic states can be found in condensed matter systems. The Majorana fermionic states in condensed matter systems, the so-called Majorana zero-energy modes (MZMs), have attracted much attention due to its non-Abelian statistics, in which the exchange operation is non-commutative [2]. Spatially separated MZMs are protected topologically, and thus provide a platform for fault-tolerant topological quantum computation based on non-Abelian braiding operations [2, 3].

Owing to the particle–hole symmetry in superconductors, $\gamma(E) = \gamma^+(-E)$, zero-energy Bogoliubov quasiparticles can be regarded as the Majorana fermionic excitation in topological superconductors. Initially, it was proposed that the Majorana fermionic states appear as low-energy excitations in fractional quantum Hall phases [4] or $^3$He superfluid [5]. There have been numerous attempts to detect Majorana fermions in topological superconductors, such as one-dimensional spinless $p$-wave superconductors [6] and two-dimensional chiral $p_x + ip_y$ superconductors [7-9]. However, the existence of $p$-wave superconductors in nature has yet to be confirmed [10]. Instead, topological superconductors have been artificially engineered to realize topological superconductivity and the MZMs via the superconducting proximity effect [11]. Topological superconductivity can be induced when a conventional $s$-wave superconductor contacts a material with strong spin-orbit coupling. Recently, diverse systems have been hypothesized [11-22] and investigated experimentally [23-35] to realize MZMs (see also detailed reviews in Refs. [36-39]).

There are two major experimental schemes to detect the MZMs in a topological superconductor. The first involves observing quantized zero-bias conductance due to the MZMs, $G = 2e^2/h$, by tunneling spectroscopy. The second requires resolving the fractional Josephson effect in a Josephson weak-link configuration. Experimental studies on quantized zero-bias conductance have claimed the existence of MZMs; however, other effects, such as disorder [20, 21] or the Kondo effect [23], cannot be ruled out. The fractional Josephson effect emerges only in topological Josephson junctions (JJs), unless a non-adiabatic transition exists between the gapped Andreev bound states (ABSs) of a conventional junction (the so-called

Landau−Zener transition), as shown by the magenta arrows in Fig. 1(a). Because an extremely high transparency is required for such a transition [30, 40], this has been reported only once in a JJ configuration to date [41]. Thus, observation of the fractional Josephson effect can be an effective scheme to confirm the MZMs in condensed matter systems.

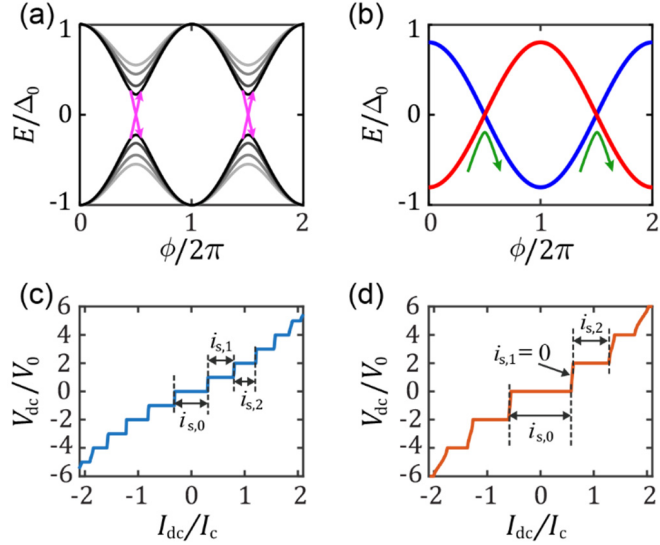

**Figure 1**. (a) Andreev bound states (ABSs) of a conventional $2\pi$-periodic Josephson junction (JJ) with transparency $\tau = 0.7, 0.8, 0.9$, and $0.95$; the higher the transparency, the smaller the gap size between two states. The magenta arrows represent the non-adiabatic transition between gapped ABSs. (b) ABSs of a topological $4\pi$-periodic JJ in the presence of the Majorana zero-energy modes (MZMs). The relaxation process, indicated by the green arrows, would change the $4\pi$-periodic ABSs to conventional $2\pi$-periodic ones. Example Shapiro steps under microwave irradiation for a JJ with $2\pi$-periodic ABSs (c) and $4\pi$-periodic ABSs (d). The direct current (DC) and voltage are normalized by the critical current $I_c$ and voltage step $V_0 = \hbar\omega_{ac}/2e$, respectively.

The conventional ABSs of a short, ballistic JJ are $2\pi$-periodic in phase difference, $\phi$, with anti-crossing at $\phi = (2n + 1)\pi$ [Fig. 1(a)]. In contrast, the ABSs of a topological JJ are $4\pi$-periodic, with an energy level crossing at $\phi = (2n + 1)\pi$ [Fig. 1(b)]. Here, $n$ is an integer number. As the energy of the ABSs changes sign at $\phi = (2n + 1)\pi$ at a topological junction, the parity of the ground state of the system also changes between even and odd at $\phi = (2n + 1)\pi$. In equilibrium, inelastic processes [represented by the green arrows in Fig. 1(b)] that violate fermion parity conservation relax the system to the ground state. This leads to $2\pi$-periodic dynamics in the current–phase relation (CPR) of a JJ, as in a conventional junction.

This also explains why Fraunhofer pattern measurements, in which the dynamical time scale of the junction is much longer than the parity relaxation time of an order of milliseconds [42-44], is not applicable for detecting the $4\pi$ periodicity of a JJ. However, fast measurements such as Shapiro steps and Josephson emission spectra, in which the characteristic sub-nanosecond time scale is shorter than the relaxation time, can be implemented for observing the fractional Josephson effect.

In this paper, we report the detailed features of Shapiro steps in the presence of $4\pi$-periodic CPR over a broad range of experimental parameters. When a JJ is irradiated by microwaves, the current–voltage curve of the junction exhibits step-like behavior, the so-called Shapiro steps, as shown in Figs. 1(c) and (d). Plateaus emerge at constant voltage steps, although the length of the plateaus (step size) can be modulated by the alternating current (AC) applied to the junction by microwave irradiation ($I_{\mathrm{ac}}$). The step height (voltage interval between these steps) is $V_0 = \frac{\hbar \omega_{\mathrm{ac}}}{2e}$ in conventional $2\pi$-periodic junctions, whereas the step height is expected to be doubled in entirely $4\pi$-periodic junctions, where $\hbar$ is Planck's constant divided by $2\pi$, $\omega_{\mathrm{ac}}$ is the angular frequency of the irradiated microwave, and $e$ is the charge of an electron. Thus, "period-doubled Shapiro steps" provide strong evidence for the presence of the $4\pi$-periodic CPR. In practice, however, the $4\pi$-periodic contribution coexists with, and is usually dominated by, the $2\pi$-periodic one, even in the presence of the MZMs, such that the step height follows $\frac{\hbar \omega_{\mathrm{ac}}}{2e}$ of the $2\pi$-periodic case. In this situation, it is expected that the steps at even multiples of $\frac{\hbar \omega_{\mathrm{ac}}}{2e}$ (even steps) become more pronounced compared to those at odd multiples of $\frac{\hbar \omega_{\mathrm{ac}}}{2e}$ (odd steps), due to the $4\pi$-periodic CPR. Therefore, to obtain information about the $4\pi$-periodic nature, the length of the plateau (step size) must be analyzed in depth.

Previous reports claiming the detection of a $4\pi$-periodic Josephson current show the distinct features of Shapiro steps. When the microwave frequency irradiating a junction is sufficiently low, the odd steps are suppressed at the low AC limit. As the AC is increased, however, odd steps emerge and their step sizes become comparable to those of even steps. In addition, the size of the odd steps for low AC is restored as the AC frequency continues to increase [29-32]. Although these features were considered to arise from the presence of a $4\pi$-periodic Josephson current, an in-depth analysis is required. For instance, the abovementioned features have not been reproduced by resistively shunted Josephson junction (RSJ) model analyses, which overlook the hysteretic behavior of a junction. To our knowledge, only two reports have considered the capacitance of a junction when describing the Shapiro steps with $4\pi$-periodic Josephson coupling [45, 46]; however, no clear explanations were provided regarding the characteristic features of Shapiro steps observed in previous measurements.

Here, we report on the step size of Shapiro steps in the presence of $4\pi$-periodic Josephson coupling. We examine the overall features of step size with respect to the broad range of representative junction parameters. The step size was obtained by numerical simulations using the resistively and capacitively shunted junction (RCSJ) model, which includes a capacitance term. The characteristic features of the Shapiro steps observed in previous experiments were reproduced only in the hysteretic junctions with finite capacitance. We also estimated the portion of $4\pi$-periodic current that exhibited good agreement with the experimental conditions. Moreover, we suggest a strategy for distinguishing a small portion of the $4\pi$-periodic Josephson current from the dominating conventional $2\pi$-periodic one, assuming realistic experimental conditions.

## II.  RCSJ MODEL

To investigate the Shapiro steps of a JJ, we use the RCSJ model [47], in which the JJ is represented as a circuit consisting of a lumped Josephson element, a resistive element, and a capacitive element connected in parallel, as shown in Fig. 2(a). The model well describes Josephson dynamics, including the Shapiro step behavior of a JJ under microwave irradiation. In this model, the current–voltage characteristics of the JJ can be obtained by solving the equations of motion of the equivalent circuit with respect to the superconducting phase difference $\phi$ across the junction, which is derived from Kirchoff's law in association with the following two Josephson relations:

$$V_J = \frac{\hbar}{2e}\frac{d\phi}{dt}, \quad (1)$$

$$I_J = I_c \sin\phi. \quad (2)$$

Here, $I_c$ is the junction critical current, and $I_J$ and $V_J$ are the current and voltage drop across the Josephson element, respectively.

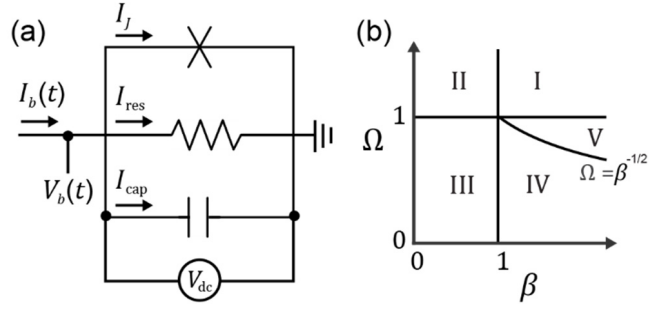

**Figure 2.** (a) Schematic diagram of the resistively and capacitively shunted junction (RCSJ) model depicting a lumped Josephson element, resistor with resistance $R$, and capacitor with capacitance $C$ connected in parallel; the bias current $I_b$ is divided into $I_J$, $I_{\text{res}}$, and $I_{\text{cap}}$ flowing through the respective components. (b) Parametric space of $\beta$ and $\Omega$, which determines the Shapiro features of a junction. The Shapiro maps in regions I, II, and V show Bessel-function behavior.

These differential equations can be solved exactly provided that the junction is under a bias voltage of the form, $V_b = V_{\text{dc}} + V_{\text{ac}} \cos \omega_{\text{ac}} t$, in which dc and ac denote direct current (DC) and AC, respectively. Here, $\omega_{\text{ac}}$ is the angular frequency of the AC bias and $V_{\text{dc}}$ ($V_{\text{ac}}$) denotes the amplitude of the DC (AC) bias voltage. It is well known that the DC through a JJ exhibits sharp spikes at quantized values of the DC voltage, $V_0 = n \frac{\hbar \omega_{\text{ac}}}{2e}$, where $n$ is an integer. The corresponding DC ranges between $\frac{V_{\text{dc}}}{R} - I_c \cdot J_n(\frac{2eV_{\text{ac}}}{\hbar \omega_{\text{ac}}})$ and $\frac{V_{\text{dc}}}{R} + I_c \cdot J_n(\frac{2eV_{\text{ac}}}{\hbar \omega_{\text{ac}}})$, depending on the initial value of $\phi$ at $t = 0$. Thus, the step size is proportional to the $n$-th-order Bessel function of applied AC voltage $V_{\text{ac}}$ (for details, see Section 1 of Supplemental Material).

However, for the proximity junction considered, the junction is current-biased following the form of Eq. (3), which leads to junction behavior contrasting with the voltage-biased case of a tunnel junction due to the nonlinear characteristics of the Josephson element.

$$I_b = I_{\text{dc}} + I_{\text{ac}} \cos \omega_{\text{ac}} t. \tag{3}$$

Combined with the Josephson relations in Eqs. (1) and (2), the total current through the JJ can be represented by a second-order differential equation with respect to $\phi$ as

$$I_{\text{dc}} + I_{\text{ac}} \cos \omega_{\text{ac}} t = I_c \sin \phi + \frac{\hbar}{2eR} \frac{d\phi}{dt} + \frac{\hbar C}{2e} \frac{d^2 \phi}{dt^2}. \tag{4}$$

This reduces to a dimensionless form

$$i_{dc} + i_{ac}\cos\Omega\tau = \sin\phi + \dot{\phi} + \beta\ddot{\phi}, \qquad (5)$$

in terms of the reduced time $\tau = \frac{2eRI_c}{\hbar}t$, reduced frequency $\Omega = \frac{hf_{ac}}{2eI_cR}$, and McCumber parameter $\beta = \frac{2eI_cR^2C}{\hbar}$, with $\dot{\phi} = \frac{d\phi}{d\tau}$ and $i_x = I_x/I_c$. Equation (5) is a nonlinear inhomogeneous differential equation that cannot be solved exactly; the solution differs based on the values of $\beta$ and $\Omega$. However, the corresponding bias voltage can be approximated as $V_b \sim V_{dc} + V_{ac}\cos(\omega_{ac}t - \delta)$ if one of the three conditions described by Eqs. (6)–(8) is satisfied, which corresponds to regions I, II, and V of the parametric space in Fig. 2(b) [48]. Under the three conditions, current flows mainly through the linear elements (the resistor or the capacitor) rather than the nonlinear lumped Josephson element.

$$i_{ac} \gg 1, \qquad (6)$$
$$\Omega \gg 1, \qquad (7)$$
$$\Omega \gg \beta^{-1/2}. \qquad (8)$$

Therefore, the Shapiro steps obey Bessel-function behavior, with an approximate solution of $\dot{\phi} = v_{dc} + \frac{i_{ac}}{\sqrt{1+\beta^2\Omega^2}}\sin\Omega\tau$, where $v_{dc}$ is a constant corresponding to $\langle\dot{\phi}\rangle = \frac{\langle V_b\rangle}{I_cR}$. In this case, the detailed Bessel-function features are determined by the values of $\beta$ and $\Omega$. The step emerges at $V_{dc} = n\frac{\hbar\omega_{ac}}{2e}$ with a step size of $I_s = 2I_c \cdot J_n\left(\frac{i_{ac}}{\tilde{\Omega}}\right)$, where $\tilde{\Omega} \equiv \Omega\sqrt{1+\beta^2\Omega^2}$ corresponds to $v_{dc} = n\Omega$ and a normalized step size of $i_{s,n} \equiv I_s/I_c = 2J_n(\frac{i_{ac}}{\tilde{\Omega}})$, as a dimensionless parameter (for details, see Section 2 of Supplemental Material).

For the remaining portion of the parametric space, the only way to investigate the Shapiro step characteristics is through the numerical calculation of Eq. (5). In this region, adjacent steps overlap as $\Omega$ becomes small. The overlapping features differ significantly depending on whether $\beta < 1$, as shown in Figs. 3(c–f). Thus, we emphasize that determination of $\beta$ and $\Omega$ is essential to investigating the characteristics of Shapiro steps; notably, most previous reports claiming the observation of $4\pi$-periodicity of the JJ CPR did not take this into account.

## III. SIMULATION RESULTS

### A. Purely $2\pi$-periodic Josephson current ($\alpha = 0$)

Before we introduce the $4\pi$-periodic phase-dependent element in the Josephson current, we must first numerically solve Eq. (5) to determine how the Shapiro maps for various points in the parametric space in the trivial $2\pi$-periodic case (see Section 3 of the Supplemental Material for details on the numerical calculation method). The representative Shapiro maps, $dv_{dc}/di_{dc}$, are given in Figs. 3(a–f) as a function of $i_{dc}$ and $i_{ac}$. The ranges of $i_{dc}$ and $i_{ac}$ are set so as to include the first few lobes of the $n$-th step for $|n| \leq 4$, to show the variation in Shapiro step behavior (more numerical simulation results can be found in Section 4 of the Supplemental Material).

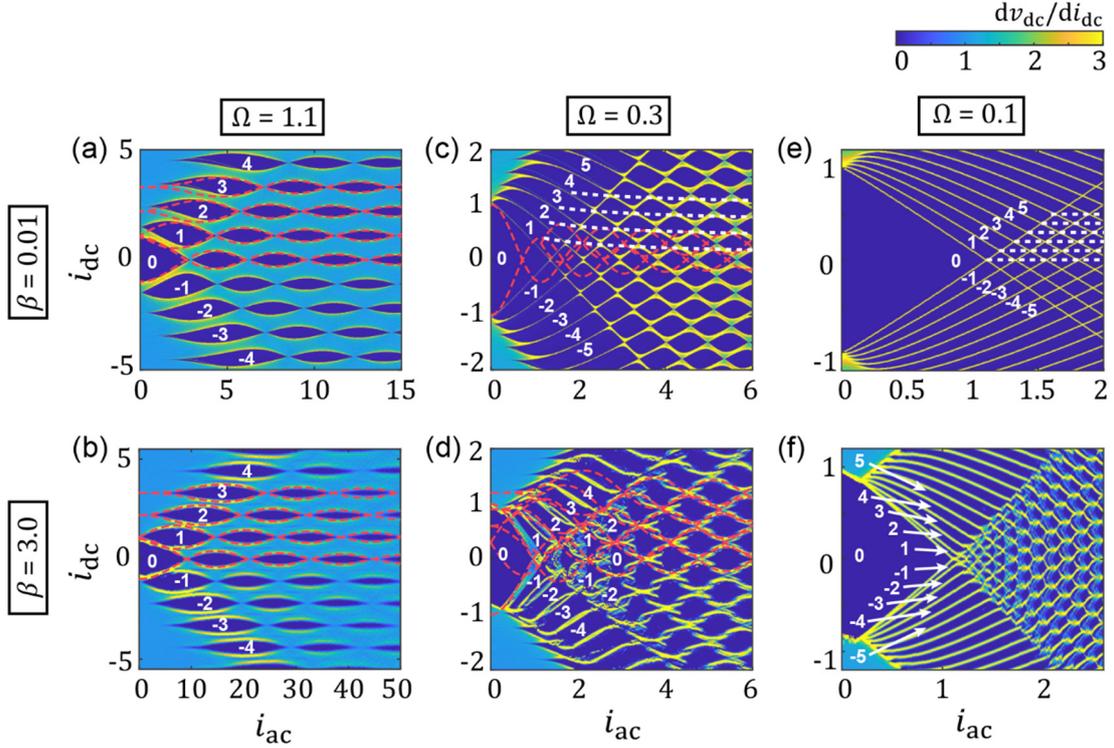

**Figure 3.** (a), (c), and (e) Dimensionless differential resistance $dv_{dc}/di_{dc}$ maps of the direct and alternating currents, $i_{dc}$ and $i_{ac}$, respectively, for the overdamped limit ($\beta = 0.01$), and (b), (d), and (f) for the underdamped limit ($\beta = 3.0$);. Here, $\Omega = 1.1$, 0.3, and 0.1 for (a) and (b), (c) and (d), and (e) and (f), respectively.

Shapiro maps for $\Omega = 1.1$ are shown for an overdamped junction in which $\beta = 0.01$

[Fig. 3(a)] and an underdamped junction where $\beta = 3.0$ [Fig. 3(b)]. Even for $\Omega = 1.1$, which does not meet the condition of Eq. (7), the calculated boundary of the steps well describes the Bessel-function behavior, $i_{dc} = n\Omega \pm J_n\left(\frac{i_{ac}}{\Omega}\right)$, denoted by the red dashed lines, for most of the lobes; this is because Eq. (6) is satisfied. In contrast, for $i_{ac} \lesssim 1$, the boundary of the steps does not follow Bessel-function behavior, as seen in the first lobes of the $n = 1, 2,$ and $3$ steps in Fig. 3(a).

For $\Omega < 1$, the Shapiro map changes significantly, as adjacent steps start to overlap. Nonetheless, Bessel features still remain with variation in $i_{ac}$ for $\Omega$ as low as ~0.3 [Figs. 3(c) and (d)]. Although the boundaries of the steps do not fit $i_{dc} = n\Omega \pm J_n(\frac{i_{ac}}{\Omega})$ (red dashed line) in fully overdamped junctions of $\beta = 0.01$, the amplitude of each lobe follows the $n\Omega + \sqrt{\frac{2}{\pi i_{ac}}}$ denoted by the white dotted lines in Fig. 3(c), which correspond to the amplitude of the Bessel function at the asymptotic limit. The effect of overlap is drastic in underdamped junctions where $\beta = 3.0$, as shown in Fig. 3(d). Although the step size does not follow Bessel-like behavior due to the overlap, sharp increases in voltage occur following Bessel-like behavior (red dashed lines). In the overlapped region, however, the emergence of fractional steps causes the voltage to be unstable.

In the fully overlapped regime, e.g., at $\Omega = 0.1$ as shown in Figs. 3(e) and (f), the Shapiro maps no longer follow Bessel-like behavior and are totally different from those of $\Omega \geq 0.3$. The step sizes of fully overdamped junctions exhibit triangular variation with the same amplitude, independent of the number of steps $n$, as shown in Fig. 3(e). In underdamped junctions, the maps become complicated. The amplitude of $i_{s,n}$ increases gradually with increasing $i_{ac}$ for more than six lobes, in clear contrast to the opposite trend for $\Omega \geq 0.3$. One important feature of this map is the unexpected suppression of the first lobe of the $n = 1$ step, marked by an arrow in Fig. 3(f). This characteristic resembles the Shapiro features of junctions with a small $4\pi$-periodic Josephson element, as described in the following.

### B. Larger portion of the $4\pi$-periodic Josephson current ($\alpha = 0.5$)

We now introduce the $4\pi$-periodic components of Josephson current by assuming that DC Josephson relation of Eq. (2) is $I_J = I_{2\pi} \sin \phi + I_{4\pi} \sin \frac{\phi}{2}$, which leads to Eq. (9) with $i_{2\pi} = I_{2\pi}/I_c$ and $i_{4\pi} = I_{4\pi}/I_c$:

$$i_{dc} + i_{ac}\cos\Omega\tau = i_{2\pi}\sin\phi + i_{4\pi}\sin\frac{\phi}{2} + \dot\phi + \beta\ddot\phi. \tag{9}$$

Here, $i_{2\pi} + i_{4\pi} > 1$ to satisfy the relation $I_c = \max\{I_J\}_\phi$. As an example, for $\alpha \equiv \frac{i_{4\pi}}{i_{2\pi}} = 0.5$ (Fig. 4), $i_{2\pi} = 0.73$ and $i_{4\pi} = 0.37$, and their sum is larger than 1.

We examine how the step size evolves for the $4\pi$-periodic Josephson component present in regions I, II, and IV of Fig. 2(a), in which the step size shows Bessel-function behavior. Whereas $\dot\phi$ is still approximated as $\dot\phi \sim v_{dc} + \frac{i_{ac}}{\sqrt{1+\beta^2\Omega^2}}\cos(\Omega\tau - \delta)$ in these regions, even with the $4\pi$-periodic component, $\langle i_J \rangle$ exhibits a distinct behavior between even and odd numbers of $n$, due to the $\sin\frac{\phi}{2}$ term of Eq. (10):

$$\sin\frac{\phi}{2} = \sum_{k=-\infty}^{\infty} (-1)^k J_k\left(\frac{i_{ac}}{2\widetilde\Omega}\right) \sin\left(\left(\frac{v_{dc}}{2} - k\Omega\right)\tau + \frac{\phi_0}{2} - k\delta\right). \tag{10}$$

The time average of $\sin\frac{\phi}{2}$ calculated from Eq. (11) is not present for odd steps. Thus, the size of odd steps $i_{s,2m-1}$ depends only on $i_{2\pi} \langle \sin\phi \rangle$, which decreases according to the ratio of $i_{2\pi}$ to $i_s = \langle \sin\phi \rangle$ for the purely $2\pi$-periodic case of $i_{4\pi} = 0$:

$$\langle \sin\frac{\phi}{2} \rangle = \begin{cases} 0 & (v_{dc} = (2m-1)\Omega) \\ J_m\left(\frac{i_{ac}}{2\widetilde\Omega}\right)\sin(\frac{\phi_0}{2} - 2m\delta). & (v_{dc} = 2m\Omega) \end{cases} \tag{11}$$

On the other hand, for $v_{dc} = 2m\Omega$, the step size is determined by the sum of $\langle \sin\frac{\phi}{2} \rangle$ and $\langle \sin\phi \rangle$, whose maximum and minimum values cannot be expressed explicitly.

Figures 4(a) and (b) show the Bessel-like Shapiro steps when the $4\pi$-periodic portion of the Josephson current becomes as large as $\alpha = 0.5$ (more results for different values of $\alpha$ can be found in Section 5 of the Supplemental Material). The black dashed lines in Fig. 4(b) represent the result of a purely $2\pi$-periodic case. The size of the odd steps clearly decreases compared to that of $2\pi$-periodic ones, as expected based on the approximation $\dot\phi \sim v_{dc} + \frac{i_{ac}}{\sqrt{1+\beta^2\Omega^2}}\cos(\Omega\tau - \delta)$ introduced above. On the other hand, the size of even steps is comparable to that of $2\pi$-periodic ones, but the minima of the lobes do not vanish completely as the two terms are, in general, out-of-phase.

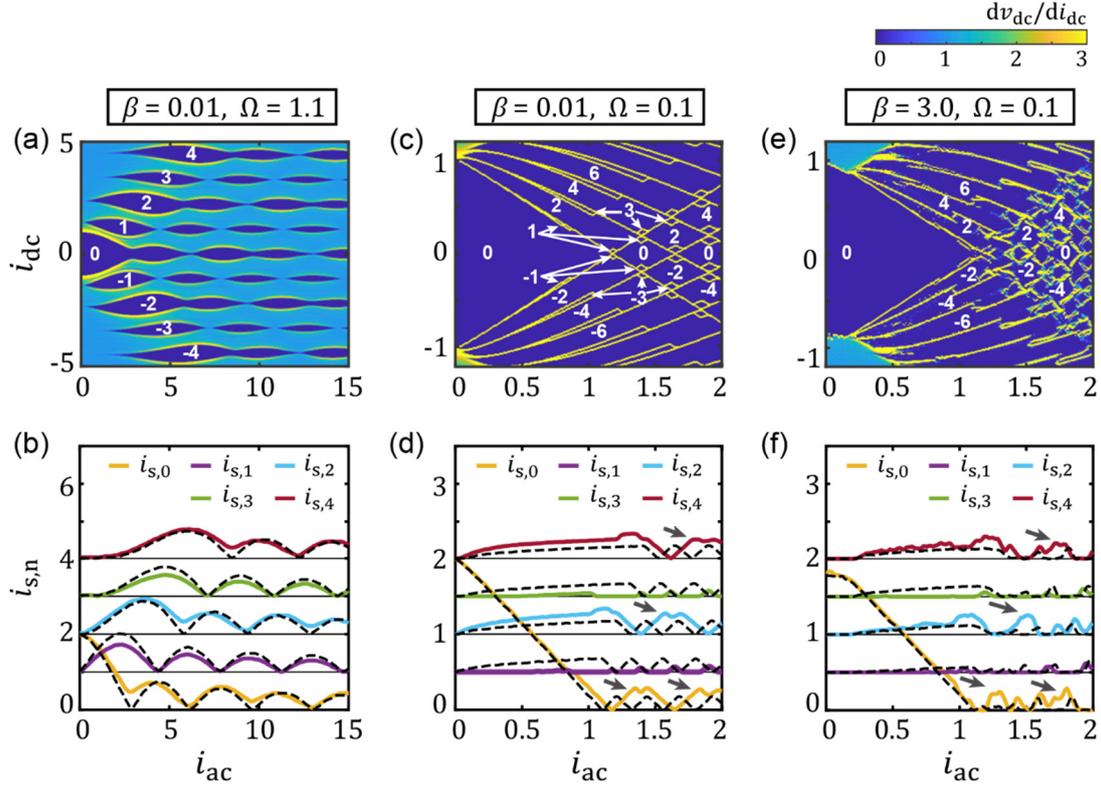

**Figure 4.** (a), (c), and (e) Representative Shapiro maps, and (b), (d), and (f) the corresponding step sizes $i_{s,n}$, in the presence of large $4\pi$-periodic components of Josephson current where $\alpha = 0.5$, for the parameter values denoted at the top of the panels: $(\beta, \Omega) = (0.01, 1.1)$, $(0.01, 0.1)$, $(3.0, 0.1)$. Each curve for $n > 0$ is offset to enhance visibility; $1.0 \times n$ for (b) and $0.5 \times n$ for (d) and (f).

The even-odd effect is more evident in the low-frequency regime with $\Omega = 0.1$, as shown in Figs. 4(c–f). The odd steps are extremely suppressed and almost vanish for most of the $i_{ac}$ region, whereas the even steps are enhanced, as shown in Figs. 4(c) and (d). Similarly, although the step size has a somewhat complicated dependence on $i_{ac}$, the enhancement of even steps and suppression of odd steps are more clearly visible in an underdamped case ($\beta = 3.0$), as shown in Figs. 4(e) and (f).

Period doubling of the Shapiro steps observed experimentally to date exhibits the suppression or absence (full suppression) of the first lobes of odd steps, whereas the even and odd steps exhibit alternating behavior in step size, similar to the purely $2\pi$-periodic case [see black dashed lines in Figs. 4(d) and (f)] for the second and higher-order lobes. However, our simulation with $\alpha = 0.5$ indicates that the enhancement of even steps is so large that adjacent even lobes merge together, as denoted by the magenta arrows in Figs. 4(d) and (f). This contradicts the observed alternating variation of step size with respect to $i_{ac}$ reported by the studies claiming the existence of 4π-periodic Shapiro steps [30-32].

C. Smaller portion of the 4π-periodic Josephson current ($\alpha \leq 0.1$)

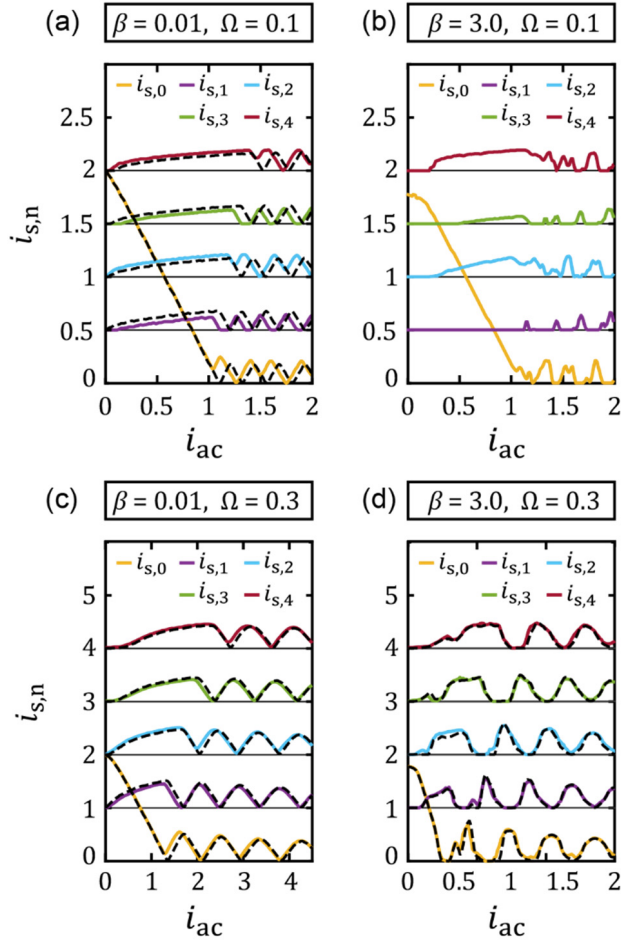

**Figure 5.** Shapiro steps as a function of $i_{ac}$ in the presence of a low 4π-periodic Josephson current, $\alpha = 0.1$, for relatively low alternating current (AC) frequency $\Omega \leq 0.3$. The values of parameters $\beta$ and $\Omega$ are shown at the top of the panels. Curves of $i_{s,n}$ for $n \geq 1$ are offset to distinguish them clearly from the adjacent one in steps of 0.5 for $\Omega = 0.1$ in (a) and (b), and steps of 1.0 for $\Omega = 0.3$ in (c) and (d). The black dashed lines represent the variation of $i_{s,n}$ for the case of 2π-periodic Josephson current only.

We now decrease the $\alpha$ to 0.1 and obtain the $i_{s,n}(i_{ac})$ for relatively low frequencies ($\Omega = 0.1$ and 0.3), as shown in Fig. 5. Although the portion of the 4π-periodic component is small, the even-odd alternation is clearly visible for the fully overlapping case of $\Omega = 0.1$ in Figs.

5(a) and (b). In the overdamped case of $\beta = 0.01$, the suppression (enhancement) of odd (even) steps seems to be insensitive to $i_{ac}$, where the first lobes of odd steps are not missing.

In the underdamped case of $\beta = 3.0$, however, the suppression (enhancement) of odd (even) steps is clear for the first lobes, where the first steps disappear completely. On the other hand, the size of higher-order lobes is not much different from that of the $2\pi$-periodic ones, with the alternation of even-odd steps being preserved. These features are similar to the experimental observations of Ref. [30], which indicate that the ratio should be $\alpha \lesssim 0.1$. They also accord with the fact that all of the junctions described in Refs. [30-32] are hysteretic with respect to their transition characteristics, which is pertinent to underdamped junctions.

As $\Omega$ increases up to 0.3, neither the first nor higher-order lobes exhibit noticeable enhancement or suppression in $i_{s,n}$, regardless of the value of $\beta$ [Figs. 5(c) and (d)]. Despite the presence of the $4\pi$-periodic Josephson element, the sizes of the Shapiro steps denoted by the solid lines of $\alpha = 0.1$ mostly follow the purely $2\pi$-periodic cases denoted by the black dashed lines of $i_{4\pi} = 0$, with slight suppression (enhancement) of odd (even) steps. This points to the possibility of survival of the $4\pi$-periodic Josephson component at relatively high-frequency AC, whereas these features were interpreted in terms of quasiparticle poisoning in previous experimental reports.

So far, we have examined how the Shapiro steps depend on various sets of $\Omega$ and $\alpha$ for representative values of $\beta = 0.01$ (overdamped) and $\beta = 3.0$ (underdamped). Here we show in detail the dependence of the Shapiro steps on $\beta$. A lower value of $\Omega$ is more advantageous to detect the small portion of $4\pi$-periodic Josephson coupling (Fig. 5). However, there is a lower bound of $\Omega$ $(= \frac{\hbar\omega_{ac}}{2e} \cdot \frac{1}{I_c R})$ with respect to observation of the Shapiro steps, depending on the junction quality and resolution of the voltage measurement. Here, we set the lower bound of $\Omega$ to 0.05 for the lowest AC frequency of $f_{ac} = 0.8$ GHz and $I_c R \sim 40$ $\mu$eV, as applied in Ref. [30].

Figure 6 represents $i_{s,n}(i_{ac})$ for $\alpha = 0.02$ and $\Omega = 0.05$, where the $\beta$ dependence of the Shapiro steps is clearly visible; here, we directly compare the variation of odd and even steps between $n = 1$ and 2, and between $n = 3$ and 4, rather than comparing the steps with purely $2\pi$-periodic ones. For the non-hysteretic junctions shown in Figs. 6(a–c), the Shapiro step behavior is almost the same as that shown in Fig. 5(a), in which all the lobes of odd (even) steps are equally suppressed (enhanced). However, the amount of suppression or enhancement is much smaller than $I_c$, and almost no change occurs in the Shapiro steps as the value of $\beta$ increases.

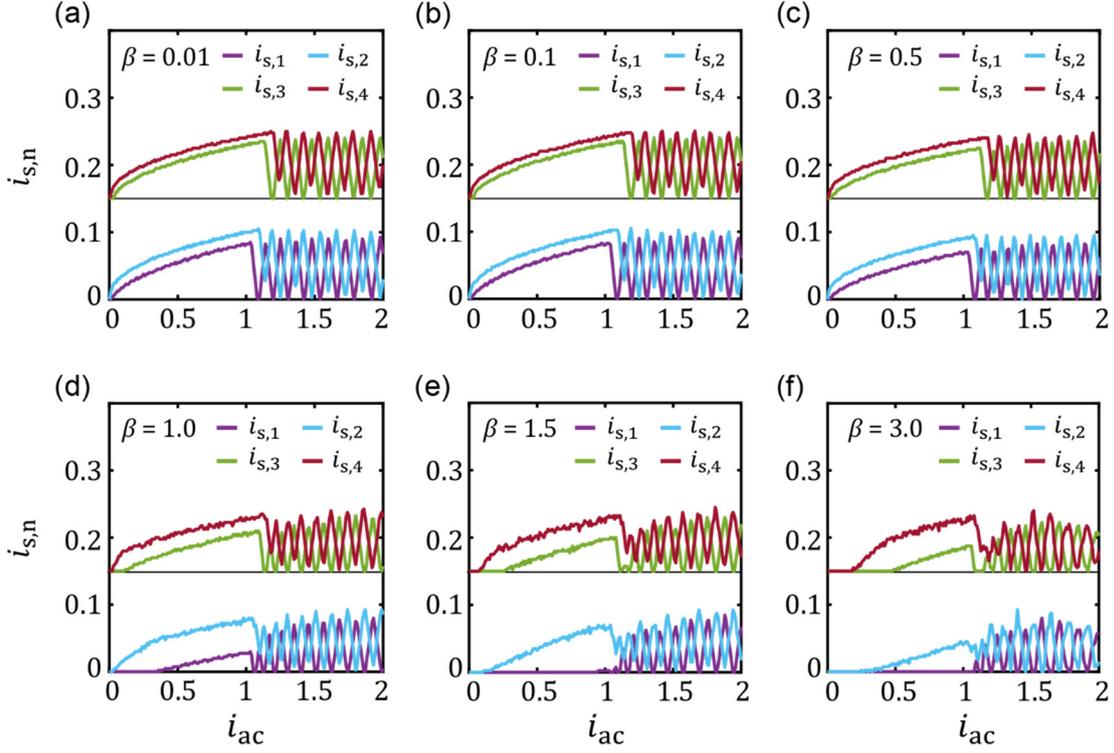

**Figure 6**. Step size $i_{s,n}$ as a function of $i_{ac}$ for different values of $\beta$ at $\alpha = 0.02$ and $\Omega = 0.05$. Curves of $i_{s,3}$ and $i_{s,4}$ are offset for visibility. (a–c) The $i_{s,n}$ for non-hysteretic junctions with $\beta = 0.01$, 0.1, and 0.5, respectively. (d–f) The $i_{s,n}$ for hysteretic junctions with $\beta = 1.0$, 1.5, and 3.0, respectively.

In the hysteretic junctions [Figs. 6(d–f)], however, suppression in the first lobe of the odd steps is clearly visible, whereas suppression in the higher-order lobes of odd steps is barely discernible. The suppression in the steps of $n = 1$ is more noticeable than that in the steps of $n = 3$, while the size of even steps remains almost the same. Also, the $\beta$-dependence of the Shapiro steps is obvious, where the amount of suppression in the first lobe of odd steps increases with $\beta$. The first lobes of the steps of $n = 1$ for $\beta = 1.5$ [Fig. 6(e)] and $\beta = 3.0$ [Fig. 6(f)] are fully suppressed, even with inclusion of the very small portion of $4\pi$-periodic Josephson current ($\alpha = 0.02$). This provides clear experimental evidence of the existence of $4\pi$-periodic CPR. Compared to the Shapiro steps in an overdamped limit, missing or full suppression of $n = 1$ steps is more clearly noticeable experimentally, despite the reduced size of the steps due to low $\Omega$ values. This suggests that an underdamped junction is more advantageous for observing $4\pi$-periodic Josephson couplings.

D. $Q_{12}$ analysis

Due to the obvious suppression in the first lobes of odd steps in the hysteretic junctions with $4\pi$-periodic CPR, the quality factor $Q_{12}$ in Refs. [30-32], specifically the ratio of the maximum values of the first lobes between the steps of $n = 1$ and $n = 2$, is a good indicator of the presence of $4\pi$-periodic Josephson coupling. Here, the value of $Q_{12}$ is estimated as a function of $\Omega$ for various parameter sets of $\alpha$ and $\beta$. We plot $Q_{12}$ for $\alpha \leq 0.15$, where the even-odd effect of Shapiro steps arising from the $4\pi$-periodic CPR is not much noticeable.

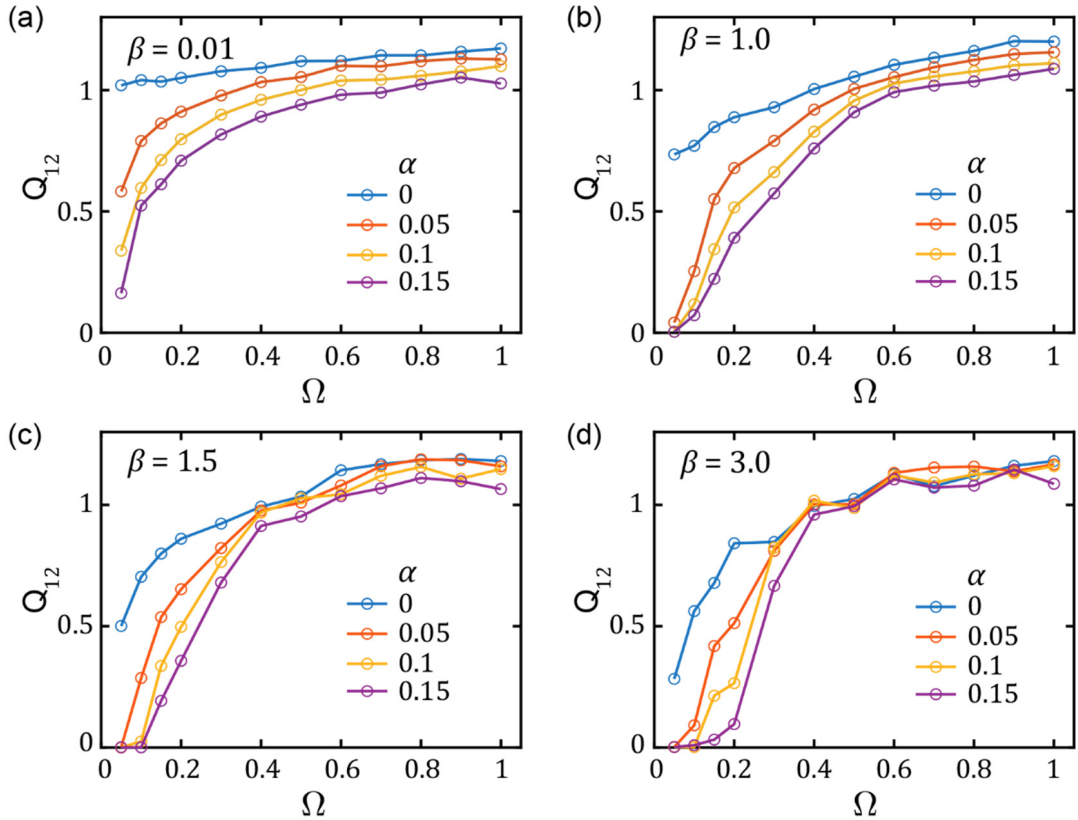

Figure 7. Quality factor $Q_{12}$: the ratio of the maximum of the first lobe of $i_{s,1}$ to the maximum of the first lobe of $i_{s,2}$, as a function of $\Omega$ for various values of $\alpha$ and $\beta$.

Figure 7(a) shows the $\Omega$ and $\alpha$ dependence of $Q_{12}$ for overdamped junctions with $\beta = 0.01$. For the purely $2\pi$-periodic case of $\alpha = 0$, $Q_{12}$ remains larger than 1, implying that the steps of $n = 1$ are larger than those of $n = 2$, regardless of the value of $\Omega$. When a non-zero

$\alpha$ is introduced, the decrease in $Q_{12}$ is almost proportional to $\alpha$. With smaller $\Omega$, suppression of odd steps is greater, such that $Q_{12}$ decays rapidly for $\Omega \leq 0.3$, as shown in Fig. 5. It is also noticeable that $Q_{12}$ does not vanish completely, even with an $\Omega$ as low as 0.05 (the lower bound of the reduced frequency) and an $\alpha$ as large as 0.15.

In contrast, when a junction is hysteretic [Figs. 7(b–d)], $Q_{12}$ can be lower than 1 for low values of $\Omega$, despite the absence of $4\pi$-periodic Josephson coupling. Especially, for $\beta = 3.0$ shown in Fig. 7(d), $Q_{12}$ for $\Omega = 0.05$ becomes as small as 0.3, which implies that $Q_{12} < 1$ does not confirm the presence of $4\pi$-periodic Josephson coupling. Thus, the values of $\beta$ and $\Omega$ are required to estimate $\alpha$ or confirm the presence of $4\pi$-periodic Josephson coupling.

For a relatively small $\alpha$, the decrease in $Q_{12}$ is so small that the $4\pi$-periodic Josephson coupling cannot easily be distinguished from the $2\pi$-periodic case in an underdamped junction, especially with a high $\Omega$ (*i.e.*, $\Omega \geq 0.3$) for $\beta = 3.0$, as shown in Fig. 7(d). However, the decrease in $Q_{12}$ becomes more evident with low $\Omega$. In this case, there is a region of $\Omega$ where $Q_{12} = 0$ depends on $\alpha$ (see details in Section 6 of the Supplemental Material). $Q_{12} = 0$ indicates that full suppression of the first lobe of the $n = 1$ step can be achieved only in the presence of the $4\pi$-periodic components of the CPR. This can be regarded as strong experimental evidence of the presence of $4\pi$-periodic Josephson coupling, even when the exact values of $\beta$ and $\Omega$ cannot be determined.

## IV. Conclusion

We introduced a strategy to design a junction confirming the presence of a $4\pi$-periodic Josephson component by analyzing Shapiro step measurements. First, it is important to increase the proportion of the $4\pi$-periodic Josephson current with respect to the $2\pi$-periodic one, which arises inevitably from the topologically trivial bulk residual modes of a junction. An increase in $I_{4\pi}/I_{2\pi}$ of only a few percent produces a large change in the Shapiro-step behavior, as shown in Fig. 7. For instance, a reduction in the channel width (area) helps to suppress the bulk residual mode for a two-dimensional (or three-dimensional) mesa structure.

Second, it is important to lower $\Omega$ $(= \frac{\hbar \omega_{ac}}{2e} \cdot \frac{1}{I_c R})$ as much as possible, as shown in Figs. 5 and 7. However, lowering $\omega_{ac}$ decreases the height of the Shapiro steps $V_0 = \frac{\hbar \omega_{ac}}{2e}$, which makes it harder to measure the voltage steps and its size within the measurement resolution. This limits the range of $\omega_{ac}$ for observing the Shapiro steps. Another way to lower $\Omega$ is to increase $I_c R$, which corresponds to the voltage of the junction at $I_{dc} = I_c$. The $I_c R$ is typically smaller than the $I_c R_N$ product of a proximity junction, due to the conductance enhancement or multiple Andreev reflection within the superconducting gap ($\Delta_0$). As the $I_c R$ and $I_c R_N$ products are generally proportional to $\Delta_0$ in the short junction limit, one can increase $\Delta_0$ to lower $\Omega$. However, when $\Delta_0$ is too large, the coherence length is excessively reduced, which weakens the Josephson coupling strength and thus reduces the $I_c R$ product for a fixed junction length. In addition, as the value of $I_c R$ depends strongly on the interfacial transparency of a junction, it is important to have sufficient superconducting interfacial contact.

Lastly, $\alpha$ is so small in usual experimental situations that a low $\Omega$ is required to observe the period-doubled Shapiro steps. However, a low $\Omega$ reduces the step size, as well as the step height, which hinders accurate estimation of the former. In this case, an underdamped junction is better than an overdamped junction for discriminating the 4π-periodic contribution. In the underdamped junction regime, the first lobes of the $n = 1$ step may be missing. Fortunately, the increase of $R$ and $I_c R$ also increases $\beta$ ($= \frac{2 e I_c R^2 C}{\hbar}$), which compliments the strategy of increasing $\alpha$ and lowering $\Omega$. If $\beta$ is not sufficiently large to observe full suppression of the $n = 1$ step, connecting a shunt capacitor to the junction can be helpful.

In summary, we examined how Shapiro steps evolve using different parameter sets of $\alpha$, $\beta$, and $\Omega$. All three parameters affect the features of Shapiro steps significantly. We succeeded in reproducing the Shapiro steps and achieved results similar to those of previous experimental reports claiming the presence of 4π-periodic Josephson coupling, with a relatively small 4π-periodic Josephson current ($\alpha \lesssim 0.1$) in a moderately underdamped junction regime ($\beta \gtrsim 1.0$). Under this condition, the $\Omega$ dependence of the Shapiro steps is also reproduced, in which the first lobe of odd steps survives an increase in $\Omega$. For an underdamped junction, the first lobe of the $n = 1$ step can be suppressed, despite the absence of 4π-periodic Josephson coupling. Therefore, we propose that full suppression of the first lobe of the $n = 1$ step must be confirmed to claim the presence of 4π-periodic Josephson coupling. To that end, we conclude that the low $\Omega$ and large $\beta$ would be advantageous to distinguish the 4π-periodic Josephson current from the trivial 2π-periodic one by observing period-doubled Shapiro steps. We expect this study to serve as a guide for investigations of the Shapiro steps relevant to Majorana-bound states in a topological JJ.

This work was supported by the National Research Foundation of Korea (NRF) through the SRC Center for Topological Matter, POSTECH, Korea (Grant No. 2018R1A5A6075964) for H.-J.L. It was also supported by NRF (Grant No. 2020M3H3A1100839) and by Samsung Science and Technology Foundation (Project No. SSTF-BA1702-05) for G.-H.L.

To whom correspondence should be addressed.

E-mail: lghman@postech.ac.kr (G.-H. Lee) and hjlee@postech.ac.kr (H.-J. Lee).